# The Laws of Nature and the Problems of Modern Cosmology

Yves Gaspar[1] · Paweł Tambor[2]

**Abstract**
The notion that nature is subject to laws is exciting from many different viewpoints. This paper is based on the context of modern cosmology. It will list the significant interdisciplinary implications generated by various aspects of the contemporary scientific discussion about the status of laws of nature, especially their dynamic nature. Recent work highlights how multiple aspects of the observed universe still lack explanation and that several problems of standard cosmology still form the object of debate. Considering these issues, several proposals have been made that entail a revision of the concept of the law of nature, according to which the nature of time and of the relation between causality and natural laws ought to be reconsidered using approaches or viewpoints which point to philosophical issues. We argue that Tim Maudlin's concept of Fundamental Law of Temporal Evolution (FLOTE) and Nancy Cartwright's notion of the nomological machine provide new insights and valuable tools that can be used in the analysis of the status of laws in the context of cosmology and of complex systems theory. The *limits* of the traditional approach to laws of nature and their mathematical formulation are highlighted in this context, as well as the fact that many of the ultimate properties of nature may turn out to be formally unpredictable or uncomputable.

✉ Paweł Tambor
pawel.tambor@kul.pl

Yves Gaspar
yves.gaspar@cantab.net

[1] Trinity Hall, Dipartimento di Matematica e Fisica, University of Cambridge (UK), Universita' Cattolica del Sacro Cuore (IT), Via Musei 41, Brescia 25121, Italy

[2] Department of Philosophy, John Paul II Catholic University of Lublin (PL), Al. Racławickie 14, Lublin 20-950, Poland

## 1 Introduction

Laws of nature are one of the fundamental subjects of philosophy of science from both the disciplinary (scientific) and interdisciplinary (ontological and epistemological) viewpoints. The conceptual schemata used in the discussion about laws are stability, necessity, and counterfactuality. Different authors name different determinants characteristic of laws. Most researchers agree that modality is a crucial property of laws of nature. This means that not every true statement, including true statements about universal general terms, is a law of nature, i.e., some statements can be true by coincidence. Another interesting property of laws from the methodological viewpoint is that they have very different exemplifications: Kepler's laws are geometric, and Newton's laws are dynamic. Nuclear physics laws (such as the laws of radioactive decay) are statistical.

Practically every methodological analysis leads to the apparent concern that not all regularities observed in nature are laws. Philosophers such as Immanuel Kant or David Hume demonstrated that generalizations created using induction are not necessarily laws in the physical sense. On the other hand, physics is developed enough as a science to realize that the only testable structures are those that contain many laws, i.e., scientific theories, rather than individual relations (Mittelstaedt and Weingartner 2005, 1). Let us list some problems implied at the starting point by discussing the laws of nature. Do laws of nature exist at all? What is the difference between a law of nature and an accidental generalization? How are laws of nature justified and validated? Is the division between dynamic and static laws the only valid one? What is the role of causality in the context of laws of nature, i.e., are they causal by necessity? Are the laws investigated by different scientific branches and disciplines (astronomy, biology, and chemistry) reducible to laws of nature?

The term *law of nature* itself can have many different understandings. Peter Mittelstaedt and Paul Weingartner suggest five meanings (Mittelstaedt and Weingartner 2005, 7–8). Law is present in the mind of its discoverer, i.e., whatever the researcher has in mind when discovering the law. This understanding does not fit a law of nature. Secondly, we speak of laws as existing in the natural world. Another insight treats law as an ideal. We only observe approximations in nature. This is the most adequate understanding of what a law of nature treats as a statement expressed in formalized scientific language. They exist in this sense – as sentences expressed in the scientific language. Laws may have different ontological statuses (existence in mind, in a sentence, or in nature). Antirealists (such as Ronald Giere or Bas van Fraassen) claim that laws of nature do not exist and provide various arguments. Some of them stem from the observation that those generalizations considered laws are, in fact, not laws. The metaphysical criticism of laws is based on the thesis that any laws concerning properties would be ontically external concerning these properties.

It is also worth reminding the reader about a philosophical (but not theological) thread involving two different understandings of the origin of laws: laws given to nature and causal powers possessed by natural objects. Hume considers laws of nature to be binding agents of the Universe. Descartes translates the laws of nature, which are tied to God's mind, onto the world of physical objects. The fundamental problem with the concept of laws of nature as directly God-given was related to their arbitrariness: to what extent was God (the lawgiver) free to shape the laws of nature?

The Aristotelian interpretation of the law, later adopted by the Middle Ages, focuses on the notion of causal powers, which are a derivative of the substantial form of a given object.

For several reasons, the category of causal powers is problematic from the ontological and epistemological viewpoints. Firstly, their categorical status demands analysis. Do causal powers result from the structural properties of a given object, or are they something more existentially fundamental? Secondly, according to the traditional interpretation, causal powers are related to a volitive disposition, a mind feature rather than a physical object. Boyle and Locke attempted to reducibly transpose the causal laws into the properties and relations between objects.

The Stanford Encyclopedia of Philosophy groups the proposed concepts of laws of nature, distinguishing, for example, approaches that reduce laws to more fundamental concepts or relationships (Carroll, 2020). First, it is necessary to mention the idea that places the law of nature in the structure of the deductive system (Lewis, 1973; 1986). A deductive system of which the law is part should maintain a balance between the informative content (the system's strength) and simplicity. The systemic approach to the laws of nature is part of a broader metaphysical context called Humean Supervenience, which can be expressed with the thesis that all facts about the world support facts about the properties and relations between the components of the world (all the facts about a world supervene on facts about which individuals instantiate which fundamental properties and relations.). The advantage of the system theory is, for example, the possibility of explaining the so-called vacuous laws, that is, general propositions about objects that remain true even when the things do not exist. Moreover, the systemic approach avoids references to available, modal, or causal concepts. A weakness is a certain arbitrariness regarding the equilibrium between the deductive system's simplicity and strength.

The approach competing with the system theory is linking the law of nature with general concepts: the law is the relation between universals. David Armstrong defines the law of nature: "Suppose it to be a law that Fs are Gs. F-ness and G-ness are taken to be universals. A certain relation, a relation of non-logical or contingent necessitation, holds between F-ness and G-ness. This state of affairs may be symbolized as 'N (F, G)'" (Armstrong, 1983, 85). The law of nature itself is not a generalization effect but a relationship between universals.

In his *Causation and Laws of Nature in Early Modern Philosophy*, Walter Ott describes the methodological boundary conditions for a diverse characterization of laws of nature, from the highly non-reductionist concept of laws governing the world to laws defined as a modal necessity occurring between the universals. According to John Carroll's idea of laws of nature (which occupies the extreme position in the top-down group of opinions), they are neither causal powers in physical objects nor a manifestation of any regularities but an explanation of these regularities. Laws are a unique type of meta-truths. This belief is antireductionist: laws cannot be reduced to objects or their properties but can be reduced to relationships between objects. Ott underlines that in Newton's opinion, gravity was not a law but a phenomenon, i.e., the result of physical bodies following a law (Ott, 2009, 8). According to the contemporary understanding, laws exist autonomously – they constitute pure facts about the physical world that do not require an external explanation.

## 2 Selected Problems in the Discussion on the Form and Status of Laws

Following Ott's intuition, in contrast to the approaches placing the law of nature in the context of a deductive system or identifying it with the relationship between universals, we propose to base the discussion on the laws of nature in the context of the problems of contemporary cosmology on two exciting concepts: (1) the concept of the nomological machine by Nancy Cartwright (Cartwright 1999) and (2) anti-reductionist and the minimalist concept of law presented by Tim Maudlin, who describes the law of nature as "ontological bedrock" (Maudlin, Tim 2007).

### 2.1 Nancy Cartwright's Nomological Machines

Nancy Cartwright is a philosopher whose works are frequently discussed in contemporary philosophy of science (Hartman, Hoefer, and Bovens 2008). Let us use some of the tools from Cartwright's philosophy of science, especially her concept of a nomological machine, to show that, once modified to suit our purposes, these tools can be applied to the standard model of modern cosmology. Initially, as is seen in *How the Laws of Physics Lie* (Cartwright, 1983), Cartwright believes that scientific models are strictly tied to a theory (theoretical models) and constitute linguistic objects (they underlie the theoretical aspect of models), or are fiction in the sense that a model is never an ideal copy of the system it represents, but rather, is a non-isomorphic description of that system. Cartwright makes a radical distinction between fundamental laws (which she believes to be fictional) and phenomenological laws, defined as the inductive generalizations of experimental results.

For instance, let us compare on cosmological grounds the interpretations of two laws with the same name but different origins: Hubble's law. The redshift–velocity–distance correlation, $zc=HL$, has the features of an experimental law (of course, it is an approximation derived from Slipher's and Hubble's observations). In turn, the velocity–distance correlation, $V=HL$, is linear and follows directly from (i.e., is the formal effect of) the cosmological principle, which states that space is homogenous and isotropic. For anti-realists such as Cartwright or van Fraassen, a logical value can only be ascribed to experimental (phenomenological) laws, which are, in essence, the knowledge about different phenomena. According to Cartwright, theoretical knowledge is fictional, false, and unrealistic. In other words, it cannot be true or false. Cartwright believes that the explanatory function of science products is essential and that simply finding a model for a given phenomenon constitutes an explanation. However, the structure of such a model follows that of a theory (because it is derived from a theory, it inherits the fictionality of the theory's fundamental laws) (Cartwright, Shomar, and Suárez 1995).

In the second development phase of Cartwright's beliefs, she proposes the concept of a toolbox of science that contains mathematical methods, models, assumptions, and many different conceptual tools. Cartwright treats models realistically in this phase as tools that represent reality. A scientific theory is neither true nor false; it is simply a means of creating a model. A theoretical model represents an empirical system, while a scientific theory is only one of many conceptual tools that aid in its construction. A theory alone is insufficient to create a model. Cartwright's evolution of views ends with the concept of models as mediators. As with the ideas of a toolbox, a theory alone is not enough to construct a model. Other factors must be considered, including measurement devices and experimental and mathematical techniques. In her first book, *How the Laws of Physics Lie*, Cartwright criticizes scientific realism, stating that the higher the explanatory power of a scientific model or theory, the lower its consistency with reality. In other words, the higher the explanatory power, the more idealized the model becomes. For instance, Cartwright accepts the realist approach concerning the objects (even unobserved ones) postulated by a scientific theory. However, in *The Dappled World*, she agrees with a mostly realist interpretation of scientific theories. She criticizes scientific fundamentalism (i.e., the belief that there exists only one set of laws that govern the universe. Thus, its opposite, anti-fundamentalism, is the idea that many scientific theories provide adequate descriptions governing reality, each of which proposes a different set of laws). Most importantly, both outlooks – fundamentalism and anti-fundamentalism – can be interpreted in a realist manner. Fundamentalists believe that every aspect of the world is governed by laws, which are the same everywhere. Anti-fundamentalists also support a positive ontology but accept that laws govern only some parts of reality and that these laws are not necessarily the same. It should be mentioned at this point that anti-realism refers not to the world's ontology but rather to the description related to that ontology.

Let us attempt to transition Cartwright's new vision of the metaphysics of reality, in which nature is governed by specific laws, onto a theoretical level. Firstly, let us apply the notion of a dappled world to a set of theoretical constructs, in this case, cosmological models. Secondly, let us interpret the SCM as a nomological machine (which, according to Cartwright, is a physical system rather than a model). Before we begin our analysis, let us explain Cartwright's physics concept's meaning of a 'dappled world.' Cartwright proposes a change in the outlook of the structure of reality and the theoretical models that represent that reality. This change, she believes, is necessary due to science's failure to create a theory of everything. In the introduction to her book, she writes (Cartwright 1999, 1): "A dappled world, a world rich in different things, with different natures, behaving in different ways. The laws that describe this world are a patchwork, not a pyramid. They do not take after the simple, elegant, and abstract structure of a system of axioms and theorems. Rather they look like – and steadfastly stick to looking like – science as we know it: apportioned into disciplines, apparently arbitrarily grown up; governing different sets of properties at different levels of abstraction."

Cartwright defines the laws of nature as descriptions of physical systems that indicate regularities related to either permanent relationships between different properties or individual causal relationships that occur regularly. It should be noted that, according to Cartwright, human knowledge is not knowledge about the laws but the nature of reality, which makes it possible to construct nomological machines (Cartwright 1999, 4). She proposes a nomological metaphysical pluralism, which she defines as follows (Cartwright 1999, 31): "Metaphysical nomological pluralism is the doctrine that nature is governed in different domains by different systems of laws not necessarily related to each other in any systematic or uniform way; by a patchwork of laws." Cartwright confronts this pluralism with scientific fundamentalism. Let us reiterate that the latter aims to discover general laws that are true and apply everywhere in the universe. Cartwright takes a moderate approach: science represents reality, modeled by a patchwork of laws specific to a given area of reality. She defines a nomological machine as follows (Cartwright 1999, 49–50): "It is a fixed (enough) arrangement of components, or factors, with stable (enough) capacities that in the right sort of stable (enough) environment will, with the repeated operation, give rise to the kind of regular behavior that we represent in our scientific laws." According to this concept, there are different types of nomological machines, each of which generates different types of laws: laws of physics, laws of economics, causal laws, or laws of probability. Importantly, ontological laws are secondary to what Cartwright calls capacities, which exist in the physical domain.

A significant difficulty in adapting the notion of a nomological machine to cosmology is that, according to Cartwright, a nomological machine is a physical system. For instance, consider the solar system. It consists of the Sun (represented by the point mass $M$), planets (represented by the point mass $m$), distance r, and forces. For the solar system to be accepted as a nomological machine that follows Kepler's laws, it cannot contain any other massive objects or factors. The presence of other planets disrupts Kepler's ideal elliptical orbits (Cartwright 1999, 52).

Another essential trait of a nomological machine about a physical system is that, on the interpretative level, it constitutes a philosophical construct, such as the different categories of determinism, laws of nature, or causal relationships. A nomological machine is an effective and idealized system. This is the case with the Earth-Sun system if we ignore the gravitational pull of any other objects. A nomological machine is a machine in that it 'produces' laws. At first glance, it seems impossible to apply Cartwright's definition to the cosmological model; for instance, because the modeled universe does not implement laws of nature in the cosmological scope, i.e., in the area of repeatable regularity. These issues will be addressed in the next section.

Before proceeding further, let us mention another possible feature of Cartwright's use of nomological machines in the context of physics and cosmology: namely, the occurrence of *uncomputable features* of natural systems. A given physical quantity is computable if there exists a recursive procedure, i.e. an algorithm, which produces, after a finite time, the n-th digit of the numerical expansion of such quantity. For example, a number like $\pi$, although being irrational and such that no algebraic equation will have it as a solution (mathematicians call $\pi$ a transcendental number for this reason), admits infinite sequences which converge towards this number such that recursive formulas can be written that effectively compute $\pi$: despite being transcendental and irrational, it is a computable number since there exists an algorithm capable of producing any digit of $\pi$ after a suitable and finite time. On the other hand, one can define numbers, such as Chaitin's $\Omega$ number (Chaitin 2004), which are uncomputable – there exists no algorithm that computes an n-th digit of $\Omega$ after a finite time. It turns out that the vast "majority" of the real numbers is actually uncomputable – the set of computable numbers is, as mathematicians sometimes say, "a negligible set of measure zero". Now, in the context of physics, one can say that if a fixed set of laws exists, which is expressed through a finite set of mathematical equations, then in principle an algorithm can be extracted from these equations that can compute a given physical property of a physical system which is considered. However, if the system behaves as a nomological machine, which can, in "fortunate" circumstances produce regular laws, then one could expect that, in other regimes or in less "fortunate" circumstances, this nomological machine will behave differently and produce intrinsically different laws that cannot be deduced through an algorithm from the previous set of laws. There would be no general algorithm that computes the behavior of physical quantities in all possible regimes or circumstances within which a nomological machine would operate. Consider a region of a cosmological space-time – viewed as a nomological machine, in a low-energy regime, this "machine" can produce regular laws such as those which are effectively consistent with observations in our low-energy cosmic environment. On the other hand, if this region of space-time is considered in a high-energy regime, for instance close the initial Big Bang cosmological singularity or close to a black hole singularity, new degrees of freedom can be excited, and the effective behavior of space-time would be different and will not correspond to the behavior which occurs far from singularities in low-energy regimes. This new behavior would appear to be uncomputable and unpredictable if one would use the "low-energy" laws or algorithms. This corresponds essentially to the approach which is proposed by Unger and Smolin (Smolin and Unger 2015) to tackle problems of modern cosmology which we view in the light of Cartwright's concepts, and which will be considered in the subsequent sections (Sect. 3 in particular).

### 2.2 The Law of Nature as an 'ontological Bedrock.'

Tim Maudlin presents his concept of the laws of nature in *The Metaphysics within Physics* (Maudlin, Tim 2007). He calls his conception modest: "My own proposal is simple: laws of nature ought to be accepted as ontologically primitive." (Maudlin, Tim 2007, 15). And then he writes: "My analysis of laws is no analysis at all. Rather I suggest we accept laws as fundamental entities in our ontology. Or, speaking at the conceptual level, the notion of law cannot be reduced to other more primitive notions" (Maudlin, Tim 2007, 18). Maudlin's modesty is also hidden in his vision of the relationship between philosophy and natural sciences. His position is expressed in the book's title: the philosophy that studies the nature of laws of nature is a derivative of physics. It must treat them as data - discovered and determined in scientific research. The researcher proposes a simple methodological assumption: metaphysics related to science can only serve as a reflection on physics.

In the explication of the logical form of the law of nature, Maudlin first refers to the form $(x) (F (x) \supset G (x))$, and finally recognizes it as a small methodologically effective in the analysis of the laws of nature. The relation between qualities F and G is necessary, but not in a logical or metaphysical sense, but a nomic sense. The nomic necessity of the law lies not in the fact that the entire expression retains its necessary character but that the property $F (x)$ necessarily implies the property $G (x)$: "It is not just that $(x) (F(x) \supset G(x))$ is necessary, it is that being an F necessitates being a G" (Maudlin, Tim 2007, 10). The researcher, however, abandons classical logical analyzes to propose a concept of the law of nature based on the analysis of the phenomenon of the temporal evolution of a physical system. Logical necessity has its rules in the laws of logic. It entails metaphysical necessity (but not vice-versa). Metaphysical necessity entails nomic (nomological or physical) necessity. Nomological necessity is grounded in the laws of nature. A good example of a metaphysical necessity is the case of the anthropic principle in its strong formulations. As expressed by Brandon Carter, "the Universe (and hence the fundamental parameters on which it depends) must be such as to admit the creation of observers within it at some stage" (Carter 1974). It is a strong metaphysical assumption about the properties of the Universe. An example of the nomological necessity in the context of cosmology is the relation between the cosmological principle and the expansion of the Universe (accepting the cosmological principle necessitates expansion).

Maudlin formulates two concepts of the laws of nature: "Fundamental Law of Temporal Evolution (FLOTE)," which is a fundamental law that describes the evolution of physical states through time, and Laws of Temporal Evolution (LOTE), which are laws or lawlike relations in other than physics branches of scientific inquiry (sociology, biology) (Maudlin, Tim 2007, 12). The description of the physical situation, and strictly the temporal evolution of the physical system, is carried out with the help of FLOTE and additional conditions: adjunct principles (for example, magnitudes of forces, the form of the Hamiltonian, or the sorts of physical states that are allowable) and boundary conditions (Maudlin, Tim 2007, 13). Laws of nature contained in FLOTE can be deterministic or irreducibly stochastic depending on the nature of the system they rule.

Two contexts are essential in identifying the laws of nature: ontological and epistemological. The first is primarily about what makes a regularity a law of nature. For Maudlin, the law of nature is fundamental (in the case of FLOTE), so nothing more fundamental is needed that a law could be reduced into (for example, the relationship between universals or the concept of necessity). From an epistemological point of view, it is difficult to discern which regularity is a law of nature. According to Maudlin, Schrödinger's equation, and Newton's laws, are examples and specific paradigms of laws in the FLOTE's sense.

In its concept of the laws of nature, Maudlin pays special attention to the relationship between laws and three categories: possibility, counterfactuals, and explanation, using which the law is explicated. The law generates a model that describes the possible behavior of a physical system in each situation, predicts its state if certain conditions occur, and explains a given physical situation.

Maudlin argues that our beliefs about the laws of nature strongly depend on three categories: possibility, counterfactuality, and explanation. In the context of modern cosmology, he gives an example of the law of gravity resulting from the General Theory of Relativity (GTR), which allows for the possibility of an open and closed universe. What it is will be shown based on empirical data, while a theory containing a specific law allows for both options. Law differs from contingent regularities in that it supports counterfactual events. This means that on an ontological level, the law makes conditional sentences concerning the likelihood of a given event occurring under a specific condition true. Maudlin claims that the truth of conditional sentences depends significantly on the status of the laws of nature: "If one wants to assign truth values to counterfactuals, then one must also accept laws among the furniture of the world" (Maudlin, Tim 2007, 7). The relationship between the laws of nature and explanation is that a law should explain a given physical phenomenon. Maudlin does not develop the theory of explanation, but such a point of view is consistent, for example, with Hempel's theory.

In the case of modern cosmology, at least the one built on the GTR, Maudlin's proposal seems to be adequate and appealing. First, the notion of the law of nature is based on a dynamic concept - the evolution of the physical system (in this case, the universe as a whole) in (cosmological) time. The GTR equations allow the definition of different classes of cosmological models. In contrast, Maudlin states, "The models can then be treated as 'possible worlds' in the usual way, and so provide truth conditions for claims about nomic possibility and necessity." (Maudlin, Tim 2007, 21). According to his concept, models are a product of laws that are a fundamental component of reality.

The author of *The Metaphysics within Physics* analyzes an example of cosmology based on General Relativity, which has problems explaining essential properties of the universe, for example, the issue of monopoles or the problem of uniformity of radiation. The GTR is an example of an effective theory that is broadly consistent with empirical data. Still, it cannot explain why the universe in the present phase of its evolution has such and no other properties. The solution within standard cosmology is the inflation model, which predicts the universe's rapid expansion stage in its early stages of evolution. Maudlin argues that we can talk about inflationary laws in the context of cosmology, which are explanatory of the standard model. "The inflationary laws render the results massively insensitive to changes in initial data and give them a high objective probability. It is fair to call this explanation subsumption under law. We can understand this explanatory power judgment if we recognize laws' role in generating models. The laws explain the data because they are obtained in most of the models generated by the laws together with the variety of possible initial conditions and results of random processes." (Maudlin, Tim 2007, 44).

## 3 Explication of the Notion of the Law of Nature in the Context of Modern Cosmology

Modern cosmology is built upon three main pillars: general relativity, the Standard Model of particle physics and the underlying structure of quantum field theory, and thermodynamics and statistical physics. The so-called Standard Hot Big Bang theory is the major theory at the heart of cosmology. This theory certainly has four great observational confirmations: the expansion of the universe, the existence and properties of the cosmic microwave background radiation, the relative abundance of light elements, including isotopes in the observable universe, and the number of counts of distant radio sources compared nearer ones. But it also raises at least ten questions or problems that remain unsolved or are still debated today (see the beginning of Sect. 3.1). We show what kind of implications of the specificity and the successes and difficulties of modern cosmology on the concept of the law of nature we observe. We observe dialectical progress in cosmology: it means tension between dealing with normal science based on empirical adequacy and speculation (anthropic reasoning, multiverse). The boundaries of our observation imply cosmology's historical and evolutional character. Let us consider the simple fact: The far and better we investigate the deep space - the later in the past, the history we get to know. So, as the time of observation goes on, we see the past, the history of the Universe. Cosmology is temporal as well as the Universe is. So not only the Universe but also the cosmology is in the process of evolution. And the evolution must be not only a matter of better observational tools and more effective instruments, but it also corresponds to the nature of the Universe where the particle horizon is moving from us with the speed of light.

### 3.1 Methodological Status of Modern Cosmology and its Problems

A few of the problems of the Standard Hot Big Bang model are the flatness, the horizon, and the smoothness problems, the singularity problem, and the nature of a quantum theory of gravity, the matter/antimatter asymmetry problem, topological defect production after GUT phase transitions in the very early universe, the general initial conditions problem and the *general fine-tuning problems* of constants of physics and cosmological parameters, the dark matter, and dark energy problems.

The so-called inflationary paradigm can address the problems of flatness, smoothness, and topological defect. Still, it introduces new problems of its own, such as new fine tun-ings, the question regarding the very identity of the "inflaton" field (the field whose energy would drive the inflationary phase), the problem posed by the proliferation of inflationary models themselves (there are more than 100 of them …).

At the heart of the questions regarding the nature of the universe in its very early stages, the so-called quantum theory of gravity would help understand the general initial conditions with which the universe started its evolution towards a remarkably homogeneous and isotropic stage in late times, far from the initial Big Bang singularity, thus producing the present universe characterized by its very special properties and values of physical constants and parameters. Regarding this, the fine-tuning problem, which corresponds to the fact that the value of constants of physics and cosmological parameters are such that even tiny variations of them could result in a different universe, is one of the most intriguing aspects – it seems that the universe behaves somewhat like a pencil with its sharp end on a flat table, exhibiting strong structural instability.

It is remarkable that, although general relativity and quantum physics are each on their own very well-tested theories till now, once one tries to unify them in a single theoretical framework, a significant number of difficulties emerge such that there is no certainty on which proposal is moving in the right direction. One of these recent high-energy physics proposals, namely superstring theory or M-theory, leads to a proliferation of possible universes ($10^{500}$ models), thereby significantly weakening the predictive power of these theories and substantially leaving the general initial conditions problem and the fine-tuning problem essentially unsolved.

A common reaction to this problem consists in invoking the idea of the multiverse model, or the so-called string landscape idea, according to which different universes, starting with distinct initial conditions close to the singularity and having distinct values of constants of physics and parameters, actually all exist (or could have existed) and form a "causally disconnected" set of worlds, of which our universe is just one "fortunate," anthropically exciting possibility, in the sense that it has suitable characteristics for the emergence of life as we know it.

Authors Lee Smolin and Roberto Mangabeira Unger (Smolin and Unger 2015) argue in their recent book that superstring theory or the multiverse/landscape model could well turn out not to be a correct solution to the crisis of modern cosmology and that a new methodology, akin to natural philosophy, could provide new ways to penetrate deeper into the structure of theoretical physics and of the laws of nature themselves, thereby outlining new possible paths towards the solution of the problems inherent to modern cosmology. New paradigms should be invoked that regard the nature of laws, the concept of time, and how mathematics is used to describe the natural world scientifically and express the inherent laws of nature.

We will consider and analyze three selected issues as necessary in speculation about the laws of nature: (1) the current discussion on the nature of the Universe and its impact on the understanding of the laws of nature; (2) the theory of dynamical systems and the notion of complexity as a useful conceptual tool in the meta-reflection on the laws; (3) the problem of the laws for the Universe as a whole dynamical structure and the status of a cosmological model as a nomological machine.

## 3.2 The Nature of the Universe and the Laws of Nature

In a general sense, one can speak of a law of physics when a set of observed natural phenomena exhibits *a form of regularity*, which stably reproduces itself at every observation, in such a way that a law of nature would manifest itself as *a recurrence of a finite set of causal connections* between states of nature. This encompasses the possibility of expressing these laws using mathematics, which enables one to *predict* the future states of the observed natural systems or to "retrodict" the past states of such systems in the case of time evolution.

Smolin and Unger argue, however, that two profound problems of modern cosmology ought to be considered:

- the first one corresponds to "false universality," present essentially in the Newtonian description of the world or paradigm and assuming that locally determined laws of nature are valid for the entire universe. This assumption is still part of general relativity and even quantum physics. Moreover, general relativity and quantum physics are assumed to hold throughout our universe and other "universes" in the multiverse/landscape scenario.
- the second one corresponds to what one could call "universal anachronism," which assumes that ideas on laws of nature pertinent only to a small part of the universe's history hold throughout the whole cosmic history. Note that these issues can be linked to Cartwright's conception of a nomological machine: considering a subset of the universe as being such a machine, there is no reason why this subset should exhibit its behavior in the same way in all spatial locations in the universe as well as at all instants of time, past and future.

To address these fundamental issues concerning the general initial conditions problem and the questions linked to fine-tuning, three essential ingredients can be considered according to Smolin and Unger, which correspond to:

- the singular universe idea, which entails that there is only one universe,
- time is fundamental and permeates everything, including the laws of nature themselves, and this would entail that, at basic levels, there would be no "fixed" set of laws of nature. Time would permeate everything, implying that sooner or later, a regular recurrence can be altered, meaning a real change of the physical direction itself.
- mathematics has excellent powers to describe the natural world, but it is also important to realize that it has immense limitations – Smolin and Unger propose a "selective realism of mathematics" and a "naturalistic" view of it.

Other fundamental aspects of laws of nature should be considered in this context. Firstly, assuming that the total number of dependent and independent variables is known (as we will see, in the case of the so-called complex systems, this is not necessarily the case), the evolution equations in cosmology and physics generally boil down to essentially a set of *fixed differential* type equations – variations of given functions are explicitly given through sets of equations linking them to the functions themselves, or other variables – this implies that there exist usually an infinity of solutions to those equations, and this renders the *choice of initial conditions*, or the justification of suitable boundary conditions inevitable.

Different approaches to quantum gravity might be able to tackle this issue in cosmology; one of the well-known proposals corresponds to the Hartle-Hawking No-Boundary proposal, according to which in the quantum gravity regime, there are no initial conditions to be fixed at the boundary of the initial space-time since there would be no boundaries. However, as mentioned before, there is no certainty regarding the validity of this Euclidean quantum gravity approach with respect to other ones.

Secondly, the explanatory models based on laws of nature usually have a *reductionist nature*: for instance, the laws of thermodynamics explain the macroscopic properties of natural systems, and these laws are interpreted in statistical and mechanistic terms using the Newtonian laws of dynamics essentially when applied to particles, atoms, or molecules, which constitute these macroscopic objects. The exact behavior of these atoms and molecules depends, however, on their microscopic properties, ultimately explained by the characteristics of elementary particles such as the quark and the electron (nucleons such as protons and neutrons are constituted by three quarks). The question naturally arises where these elementary particles, such as the electron or the quark, get their properties from. By properties, in this case, we mean mass, electric charge, the color charge of quarks within quantum chromodynamics, and the spin quantum number (fermions or bosons).

The Standard Model of particle physics does not explain these parameters. A reductionist attempt that tries to go beyond the Standard model of particle physics is superstring theory or M-theory. In these theories, elementary particles correspond essentially to vibration modes in a higher dimensional space-time of extended objects such as one-dimensional strings – the whole known particle spectrum should be produced in this way. Still, a very special kind of topology of compactified higher dimensions (such as the Calabi-Yau manifolds) is necessary. Compactification entails that the extra space dimensions are "curled up" on a very small scale since they ought not to be observable at ordinary energy levels at the so-called Planck scale. However, one is then confronted with the problem of explaining why extra-dimensional space would have this very special topology necessary to produce at least the known particles of the Standard Model: where does hyperspace get its topology from? At present, no convincing compactification mechanism is known that produces these topologies and, most importantly, guarantees their stability: why would only three space dimensions be extended and macroscopic while the other six extra space dimensions of the theory remain compactified on a very small scale?

One should then also address the complicated problem posed by the enormous number of degrees of freedom present in solutions to complex field equations, which are, at least, characterized by instability. Instead of simplification, one is confronted with increasing complexification. One sees that the reductionist approach severely limits the ultimate explanatory power of theoretical physics models – one might argue that new, more profound theories will be found. Still, the latter will introduce new constants, parameters, and fine-tunings. By the way, cosmological inflation also provides such an example: indeed, some problems and fine-tunings of the Standard Hot Big Bang model are addressed, but inflation introduces new questions, parameters, and fine-tunings of its own, and one has a proliferation of inflationary models which offers many different models as a plausible explanation without providing criteria to be able to choose among a large set of scenarios.

Another fundamental aspect on which Smolin and Unger base their argument corresponds to re-examining the nature of laws of nature and analyzing the relationship *between laws and causality*. According to this perspective, *causal connections ought to be* *considered fundamental and primitive*: such an idea of causal connection entails that they form essential features of nature linked to the concept of time and that they are not just "mental constructs" or abstract explanatory inventions. A state of nature at some instant in time would determine possible future states of nature according to this primitive causal connection. According to this viewpoint, a law of nature would manifest as a recurrence of a finite set of causal connections between states of nature. Time permeates everything and is real, implying that sooner or later, a regular recurrence can be altered, which would entail a fundamental change of the natural law itself. Note that the way a law would change can also be subject to change, such that no universal computation would allow one to predict how time would affect the laws of nature in all circumstances.

In the traditional viewpoint, causal connections are cases of general fixed laws of nature. Within the perspective suggested by Smolin and Unger, laws of nature manifest themselves as specific forms of causal connections exhibiting regularity and recurrent behavior. Moreover, in the new proposal of Smolin and Unger, there is no "separation" between fixed laws and initial conditions since the laws themselves are subject to change. In this way, a new viewpoint can offer new possibilities to tackle cosmology's initial conditions and fine-tuning problems.

To speak about a "meta-law" that would govern the change of the laws of nature, Smolin and Unger consider the possibility of envisaging the unification between states of nature and the laws of (Smolin and Unger 2015, 476–79):

> ...the distinction of laws and states breaks down. This new proposal is also realized in a simple matrix model. Instead of timeless law determining evolution on a timeless space of states, we have a single evolution that cannot be precisely broken down into law and state. Formally, what this means is to embed the configuration space of states and the landscape-parametrizing laws into a single meta-configuration space.

Furthermore,

> in this model of a meta-theory, the meta-state is captured in a large matrix, X, which we take to be anti-symmetric and valued in the integers. .... The meta-law is a simple algorithm that yields a sequence of matrices $X_n$, ...

Moreover,

> ...the question of 'Why these laws' becomes subsumed into the question of 'Why these initial conditions' in a meta-theory.

This would still not solve the problem of explaining the values of cosmological parameters and constants of the Standard model. However, it could provide a new methodology and strategy to find a possible answer.

As far as possible limitations to the explanatory power of laws of nature in the general context of cosmology are concerned, it is also interesting to note that several recent treatises mention the problem posed by the "qualia" of things or by the perceived properties of natural systems in the world by conscious observers. First, Roderich Tumulka in his attempt to build a purely mathematical model of the universe, explains how the conscious perception of qualia implies uncomputable aspects, considering the red color (Tumulka, 2017). Tumulka argues that no mathematical computation can explain or have as an output *how red appears* to a conscious observer. According to modern science, one can imagine teaching everything known about the laws of electromagnetic radiation and the neurosciences to a person to make them understand what red light is. However, if this person has never actually observed red light, he will be unable to know *how red looks to him*. Even when experiencing red light, we have difficulty answering the question, "how does red light look to you?". This might hold as well for any perceivable property of any natural system. In this context, alternative viewpoints exist regarding qualia, and which correspond to the so-called "externalist" approaches to mental contents in general. According to such views, proposed amongst others by Dretske (Dretske 1996), perceptions are explained through physical elements that happen "outside the mind", within the "external" physical or natural world. Viewed in such a way, qualia would be explicable through physical laws and be ultimately computable and predictable. The general question is still open, and authors such as Tumulka argue that the existence of qualia poses serious difficulties in depicting or modeling the universe as a purely mathematical structure. The existence of conscious observers would seriously challenge such a view.

Secondly, Smolin, in the cited book, introduces speculative proposals regarding qualia that might have implications for the philosophy of the mind (Smolin and Unger 2015, 480–83). His argument essentially consists of saying that qualia correspond to events in the universe that have no precedents and invoke genuinely novel events that *do not follow a fixed set of laws*. One of the analogies considered corresponds to the fact that habitual human actions are unconscious in people –. In contrast, conscious perceptions and the correlated events in the brain would entail novel events. Both perspectives imply that qualia lead to uncomputable aspects, to properties that cannot ultimately be the result or the output of a recursive or algorithmic way of thinking. Also in this context, an "externalist" approach to mental states and qualia would yield different conclusions, since qualia would be explained through physical factors "external" to the mind and therefore be entirely computable and predictable. It is remarkable that such issues, so close to the traditional domain of philosophy and still forming the object of debate, emerge in profound and modern treatises on theoretical physics and cosmology.

As a different example of mathematical models to understand the new relationship between a causal connection and changing laws of nature, one can consider the general properties of complex dynamical systems, which would show how very complicated behaviors can emerge from the action of straightforward basic rules.

### 3.3 Complexity and Nature's Laws

Chaotic behavior occurs in a wide variety of physical systems. A familiar description of an essential feature of chaos in meteorology corresponds to the *butterfly effect*: a butterfly flapping its wings in Milan can cause tornadoes in Oklahoma (and vice-versa), thereby exhibiting *strong sensitivity to initial conditions* of the system. Suppose one studies the evolution of a chaotic system starting from two very nearby sets of initial conditions. In that case, the future evolution emanating from the two sets will differ exponentially with time, causing thereby quantitative errors in the prediction of the future states, which manifest themselves, for instance, in a weather forecast, because of small (and inevitable) numerical errors made by computers lead to significant errors in the description of the late-time state.

In Newtonian gravity, the three-body problem is known to exhibit chaos. Spin-orbit couplings between particles, accelerator beams, and similar systems can also display chaotic behavior. The equations of motion for such systems cannot be integrated, and no general formula exists which gives their state at all times. Physicists study various approximations and simulations of these systems, several *statistical or scaling properties*, and *escape rates* can be determined. *Other average quantities can be computed because the trajectories characterizing the evolution of these systems are essentially random*.

However, some dynamical systems exhibit a *degree of complexity that lies beyond chaos*. As is studied by Moore (Moore, 1990), for very simple dynamical systems, the computations of average quantities are impossible: their dynamics is not only random. It exhibits an additional *degree of complexity*. In the work of Moore, an example is constructed of a three-dimensional potential in which the motion of a single particle possesses this "new" degree of complexity related to the undecidability (impossibility to determine whether an assertion can be proved) of some system features. For such complex systems, the basins of attraction are not merely recursive. The undecidability regarding characteristics of the dynamics causes these sets to be more complex than fractal sets: at every magnification scale, qualitatively new behavior shows up. Even if the initial conditions were known exactly, the system's evolution would still be unpredictable because of this *additional degree of complexity which superposes itself onto the chaoticity of the system*. One could argue that such systems can be understood by comparing them with *Turing machines*.

The following definition can characterize another instance of such dynamical structure: consider the *class* of systems whose number of variables is a variable of the system. Natural ecosystems correspond to a complex structure in the following sense: they are generally composed of many individual units having their own possibly chaotic evolution and a hierarchical structure that characterizes them. The different parts of the hierarchical structure can interact in complicated ways (interconnectivity), which enriches the dynamics further.

As an interesting mathematical toy model, and as a general theoretical toy model for a nomological machine in Cartwright's sense, one can consider a set of four reflecting spheres in space located at the edges of a square, and consider the observable reflection patterns in each sphere. Light rays will travel between the spheres, and each of them will be reflected into each other – *this would correspond to a simple primitive causal connection considered in our discussion* - each sphere will contain the image of all the other spheres. In this way, an observer can perceive a sequence of nested spheres (or circles) within each sphere which can be infinite in principle. One can imagine that each sphere corresponds to a part of the natural environment. Suppose one labels the spheres by letters such as A (for the lithosphere), B (for the atmosphere), b (for the hydrosphere), and a (for the biosphere). In that case, the observed sequence of "interactions" between the spheres corresponds to a series of nested circles, each labeled by a specific letter. For instance, the observer could perceive in the sphere b either the reflection of b (itself), of a, of B and A. This leads to four "two lettered words," bb, ba, bB and bA. Within each of these secondary "circles," subsequent reflections or disks can be observed: bbb, bba, bbB, bbA or bab, baa, baB, baA or bBb, bBa, bBB, bBA or bAb, bAa, bAB and bAA. As long as the reflection process continues the observable sequences tend to reach the infinite length and correspond to "infinite words" like bBaBbbaBAAbaAbaa ….

In the book by Mumford, Series, and Wright, the *infinite word trees* corresponding to the reflection between spheres are studied, and deep links exist with mathematical concepts such as the Schottky and the modular group structure used to represent reflections (Mumford, Series, and Wright 2002). Also, *infinite words with no periodic pattern are* analogous to the infinite decimal expansion of irrational and transcendental real numbers. Transcendental real numbers, like π, cannot be a solution to any algebraic polynomial-type equation. Algebraic equations can describe only approximations or "truncations" of these numbers. *Some patterns can be periodic or exhibit recurrent behavior, leading to a regular law of nature; other* patterns would be non-periodic and correspond to a change to those fixed regular laws.

There is a possible relation between the patterns generated by the reflections of the spheres and the above definition of complex systems, which can have a number of variables which itself is variable. Indeed, define a variable as an *elementary recursive cell* within some program that represents an *algorithmic compression* of a given phenomenon or sequence of events (Chaitin and Doria 2012). One could consider physics and even science as the search for algorithmic compressions: a set of natural phenomena can be "compressed" if one finds a program that generates the observed sequences, such that the "length" of the program is shorter than the length of the sequence itself. A mathematical equation is an example of an algorithmic compression: a polynomial equation enables one to obtain, in most cases, a recursive formula, such as continued fractions and expansions, that generates the number, and which is a solution to that equation. Since transcendental numbers can be described only piecewise by polynomial equations, the system which would correspond to this sequence would be characterized by several variables, which is itself variable: each time a new number of decimals or of "word letters" are added, a new polynomial equation has to be considered to describe the associated number, and thus a new recursive cell variable has to be introduced in the program.

In the context of the evolution of natural ecosystems, the changeability of the number of variables would correspond to the possibility of emergent variables and the change of the laws of nature. As far as the above mathematical model of reflecting spheres is concerned, one ought to note the following:

– The nature of the sequences depends critically on the position of the observer in the system and on the fact that the observer itself can intercept and halt some reflection processes: thus, the act of observation can alter the sequential patterns themselves. Also, the distance between the spheres influences the sequence of reflections that can be observed. These features correspond to what one would expect from a nomological machine in Cartwright's sense: no fixed, general patterns are produced, the observed reflections in a given sphere depend on the configuration of the system and are situation-specific.
– Although chaoticity is a feature of the system, its complexity corresponds to higher levels induced by its hierarchical structure: one could argue that the system can be compared to a Turing machine.
– This nomological machine would produce regularities or natural laws in two ways. First, *sequences of nested spheres could exhibit some recurrence* – in this case, a general recurrence formula can be found, yielding each term of the sequence. Second, the *existence of a so-called specific, non-arbitrary limit set*, to which the sequences of all nested spheres converge, because of Cantor's theorem on nested compact sets, corresponds to a stable feature of the late-time dynamics and would thus reflect specificity linked to convergence, as well as non-arbitrariness. Although the sequences can exhibit randomicity, they converge at infinity to a well-defined limit. An observer in such a world would have to "stand back" and not be part of the process to perceive the ultimate limit set.

### 3.4 The Problem of the Law (Laws) for the Universe

According to Smolin and Unger, the mutability, and the stability of the laws of nature may be characteristic features of the universe's history. The same processes that give rise to stability may also produce mutability - this mutability ceases to contradict their stability once both are placed in the context of a historical view. In a unique cosmological context, a single history entails that at high energies, where many degrees of freedom can be excited, such as in the neighborhood of the initial singularity or the vicinity of black hole singularities, the mutability of the laws manifests itself, thereby enabling the change of the laws of nature. Within the same unique universe and through its single history, regions characterized by low energies, where fewer degrees of freedom are excited, the effective laws of nature manifest stability and regularity. The same general process, underpinned by the concept of time-like causal connection (linked to necessity?), simultaneously gives rise to changeability and stability (both linked to contingency?).

Within the context of complex natural systems, such as the earlier mentioned toy model inspired by the environmental sciences and consisting of a set of reflecting spheres, stability manifests itself in two possible ways: firstly, the existence of convergence and of the associated non-empty limit sets, regardless of the sequences which can be regular or not, is a stable feature. Secondly, through the causal connections generated by the reciprocal reflections, recurrent patterns of reflections can exist, as well as unexpected emergent patterns generated, however, by the same causal process within the same system, i.e., the reflections. The observable mutability would depend on the position of an observer with respect to the spheres or on the relative distance between the spheres. Also, in this case, the process simultaneously gives rise to changeability and stability.

As far as the Lemaitre-Hubble law is concerned, which describes the expansion of the universe, a question that one could pose is whether this law is linked to necessity or contingency. One could argue that a law is necessary if it is necessary in every possible world. It turns out that general relativity and quantum physics are assumed to hold even beyond standard cosmology, i.e., even in the multiverse or many-world scenarios. In addition, the expansion law of our universe defines the history of our observable universe and, in this sense, is deeply related to the possibility mentioned above of generating high-energy and low-energy cosmic epochs (high energy at early times, low energies at late times in cosmic expansion).

The standard cosmological model (SCM) is called a *concordance model*. The SCM also approximates reality accepted by the cosmological community and interprets physical facts within this paradigm. In other words, this is an extremely interesting situation in terms of methodology. From the viewpoint of empirical adequacy and predictive power (primarily in retrognosis), SCM is an effective theory (it provides a consistent and non-contradictory description of the world on the cosmological scale). To use Karl Raimund Popper's terminology, SCM is a well-corroborated theory.

Conversely, from the viewpoint of cognitive insight into the nature of dark energy and dark matter, cosmology uses the notions of fluid and perfect fluid and their density and pressure, constituting a phenomenological description. For this reason, and thanks to the uniqueness of its subject of study, cosmology remains a common ground for philosophical speculations, theories, and observations. This is also why cosmology is such an exciting field of research on the interaction between science and philosophy and on explanatory models, which in cosmology take on an entirely new form. From the methodological viewpoint, the situation in modern cosmology is very unusual. On the one hand, it is an undisputable achievement that we now have a model at our disposal that is adequate, descriptively and predictively, and even simple. On the other hand, the situation can be interpreted as a crisis – primarily because SCM uses the notions of dark energy and dark matter but never explains what they are. The fact is that despite cosmology's constant development (especially concerning observational potential), the Lambda Cold Dark Matter has been the model that best describes the available data for 20 years. We want to understand why the alternative models continue to lose against SCM. We propose this because the $\Lambda$CDM is structurally stable, the simplest, and flexibly adaptable to the data.

The most interesting issue for us about cosmological models, especially SCM, is their relationship and functions regarding scientific theories and data. This paper mentions problems concerning the definition of law in cosmology, which prevent scientists from explaining the law of the Universe by simply deducing it from a theory. It should be reiterated that scientific theories in physics usually explain laws, which in turn are applied to repeating phenomenological instances within a physical system. Cosmology, rather than explaining laws for the entire Universe, concerns itself with constructing a model, i.e., a representation of the Universe, to provide a consistent and relatively complete description of what our Universe is like.

A cosmological model – a model of the Universe – is not a model for theory because it constitutes part of a scientific theory that allows for a better understanding of the said theory. A cosmological model creates a model of the Universe, i.e., a model constructed to ease our understanding, rather than the notions within a theory. Indeed, in the relationship between GTR and the cosmological model, the latter is not considered a model for the former, but rather, GTR is a theory for the cosmological model. Theory serves the cosmological model in the sense that it provides crucial theoretical tools used to construct a model of the Universe. Consequently, in the theoretical domain, it is not the theory that is the goal but the model of the Universe. The construction and testing of cosmological models rely heavily on a synergy between observational and theoretical elements, which complement each other and are incomplete when separated from each other. This is because the modelled Universe, on the one hand, is considered unique, and on the other, it is only partially available for observation.

According to Cartwright, the relationship between a theoretical model and reality is analogous: a model is an analog of a phenomenological law, i.e., it is like an experimental law. Cartwright uses the term 'simulacrum' to indicate that, in a sense, such models pretend to be reality. Of course, it is easy to prove the discrepancy in this concept and conclude that models can be related to reality instead. Let us elaborate on this thought in the context of cosmology by showing that a model is like a given phenomenon because it represents some features of that phenomenon. This view resembles the one held by Giere, who believes that the primary purpose of models in science is to represent reality (Giere, 1988, 743).

In what sense can a cosmological model function as a nomological machine? This is due to its reference to the universe as very specific and different from the other models used in physics. A cosmological model is not a model in the same sense as a planetary one because it does not represent any specific physical system that functions as a part of a greater whole; instead, it is the model of everything that physically exists. In cognitive terms, a cosmological model is bound with the universe, which humanity cannot learn about other than through a model. Furthermore, a cosmological model represents the universe more than the theory it was built on.

Despite these reservations, we can present arguments that cosmology, which investigates the only available universe, can formulate laws of nature in the typical sense. Using the predicate 'for each x, f(x) implies g(x)' is permissible because the cosmological principle stipulates that space is homogenous and isotropic. From a mathematical standpoint, this means that space has a constant positive curvature with the R–W metric. Thus, the cosmological model is a nomological machine that creates the law of the universe's expansion.

Let us reiterate that cosmology investigates the universe on the largest possible scale, where any regularities are referred to as the laws of cosmology. We will show that the expansion of the universe is one such law. Consider the following parametrization of the metric: (Baryshev, 2006, 25)

$$ds^2 = dt^2 - a(t)^2 d\chi^2 - a(t)^2 I_k(\chi)^2 \left(d\theta^2 + sin^2\theta d\varphi^2\right)$$

where: $\chi$, $\theta$, $\varphi$ are the comoving 'spherical' coordinates $I_k(\chi) = sin(\chi)$, $\chi$, $sinh(\chi)$ for a closed, flat universe and an open universe, respectively, and a(t) is the scale factor. In expanding space, the metric proper distance r between the observer and a galaxy with a stable position given by equals $r = a(t)\chi$.

Let us define the relative velocity at which an object (located at the position of another object with the coordinate $\chi = 0$ on the surface (t=const)) is moving away as:

$$V_{exp} = \frac{dr}{dt} = \chi \frac{da}{dt} = \chi \left(\frac{da}{dt}/a\right) a = H(t) r$$

The above equation has been derived directly from the metric, which follows from the cosmological principle. In turn, the mathematical formulation of the cosmological principle is the R–W metric and is interpreted as follows: the relative velocity at which objects move away is proportional to their relative distance r. The proportionality coefficient is the Hubble parameter, which becomes a function of the cosmological time t.

Yuriy Baryshev addresses a critical issue: what does the universe's expansion mean from a physical standpoint? In other words, he asks about interpreting the law of expansion. The law is universal in that it concerns all points in space that are not the basepoints of that space by the cosmological principle. We may think of the law of expansion as preceded by the universal quantifier 'for any objects located at points […]'. The universe's expansion means that the space between any two points is growing. Baryshev (Baryshev, 2006, 26–27) writes: "The Real Universe is not homogeneous, it contains atoms, planets, stars, galaxies. Bondi (1947) considered spherical inhomogeneities in the framework of GR and showed that inside them, the space expands slowly. Bounded physical objects like particles, atoms, stars, and galaxies do not expand. So, inside these objects, there is no space creation. Therefore, the creation of space is a new cosmological phenomenon, which is not and cannot be tested in a laboratory because the Earth, the Solar System, and the Galaxy do not expand".

Consequently, the law of expansion has no analog in physics. It implies that the universe does not follow the law of the conservation of energy. Baryshev provides a good insight into the breaking of the law of the conservation of energy as a fundamental principle in physics: (Baryshev, 2006, 27) (see also (Harrison, 2000, 276)): "[…] that in a homogeneous unbounded expanding FLRW model one may imagine the whole universe partitioned into macroscopic cells, each of comoving, and all having contents in identical states. The -$p\,dV$ energy lost from any one cell cannot reappear in neighboring cells because all cells experience identical losses. So, the usual idea of an expanding cell performing work on its surroundings cannot apply in this case."

Our analyses of the law of the expansion of the universe lead to the following conclusions:

- The Hubble-Lemaître's law is global, i.e., it concerns the dynamic behavior of global spacetime.
- According to one interpretation, space is created because of expansion. Thus, the amount of space between any two points increases due to this expansion.
- The law of conservation of energy as a fundamental principle in physics is broken. This is the only case in physics, as it is challenging to imagine physics without this law.
- Equations of the universe's evolution (Friedmann's equations) can be reduced to equations that describe a simple mechanical system: the movement of a point molecule through a barrier of potential.

The mechanism of Cartwright's nomological machine is physical and generates a given law (a law of physics, economics, etc.). This is precisely the mechanism of the universe's expansion in the context of cosmology. The evolution of the universe is entirely contained in the relationship, $H(z)$, between the Hubble parameter and redshift. This relationship can also be shown to be the only non-trivial piece of information encoded within the metric of the spacetime (Padmanabhan, 2003, 243).

In her newest book, Nature, the Artful Modeler, Cartwright expands upon her concept of nature and the causal powers and laws that exist within nature. She lists three interpretations of nature in the context of its nomological structure. Firstly, nature follows laws as if it follows its behavior textbook. Secondly, nature follows a 'habit' based on its past behaviors. Both concepts entail accepting some form of determinism or principle of causality. Thirdly, and this is the approach Cartwright herself leans towards, nature is 'fickle': it does not follow any specific past habit (Cartwright 2019, 5). If we substitute the word 'nature' with 'universe,' we find that the latter is not 'fickle' (to use anthropomorphism) but that its evolution is unique in that it does not replicate any habit. Instead, it is a one-time-only event and is reconstructed as such by its model. Under Cartwright's definition, a cosmological model can be called a situation-specific model, which means that we treat it as neither a construct obtained from a theory through deduction nor a result of the laws of nature. Knowledge about the universe acquired through a model does not strictly constitute propositional knowledge. It is 'knowledge how' rather than 'knowledge that.' Of course, this is not in the sense of an acquired skill but the purpose of the exposition of the law of dynamics in the modeled universe. The explanation explicitly provided by the SCM does not concern the nature of the universe's constituents, but rather, it is a functional explanation of how the universe evolves. According to Cartwright's model, the main practical question in constructing and testing a cosmological model is not, 'Are the scientific rules and laws true?', but instead, 'How does our know-how about the models produce valuable knowledge?' (Cartwright 2019, 22). The dynamic nature of the universe translates into the dynamic nature of how we learn about it. The effectiveness of the Bayesian methodology, which is used in cosmology, also confirms this evolutionary character of humanity's knowledge about the universe.

## 4 Conclusion

The presented discussion of laws exposed them to a debate on their static or dynamic status. A first conclusive remark concerns the contingent nature of the laws compared to the laws of logic being genuine and necessary. The argument favoring the laws of nature as natural laws is simply the result of their form and empirical verification. The conclusion is that all laws of nature are general and have been strongly confirmed through experience in dynamic and static conditions. Therefore, general laws and strongly confirmed laws constitute strict laws.

Mittelstaedt and Weingartner list eight interpretations of what natural laws are. Essentially, they describe the organization of objects, properties, and relations in each physical domain. The condition of a given range of effectiveness applies. What type of effectiveness precisely? The applicability of law should be its internal property. Within a relevant set of assumptions, laws manifest their limitations. Natural laws describe the properties maintained and invariant within a given physical domain but with no connection to the objects themselves; in other words, laws apply to certain morphisms. Natural laws are invariant within a particular set of parameter changes. An interesting question arises: Can structural stability property be used here for the laws of nature? Laws are predominantly invariant within spacetime. One may also consider the temporal permanence or variability of laws. Overall, laws apply in all or most cases. Proper laws are a good approximation of natural laws if they are strongly confirmed and contain certain informative content. Appropriate laws describe objective reality and are a fundamental part of scientific theories.

The second exciting conclusion places the discussion about the laws of nature in the context of the debate between realism and antirealism in the philosophy of science. John Carrol asserts that the realistic position dominates among scientists and philosophers of science (Carroll, 2020), but there are significant proposals that promote well-defined and solid antirealistic interpretations of the ontological status of laws (Giere 1999; van Fraassen 1989). Nancy Cartwright is an example of a researcher who is difficult to be classified as a realist or antirealist. However, she proposed a very interesting meta-theory of laws in physics. She, for example, distinguished between true laws and those that are general. "To grant that a law is true – even a law of 'basic' physics or a law about the so-called 'fundamental particles' – is far from admitting that it is universal -that it holds everywhere and governs in all domains" (Cartwright 1999). According to her, actual laws-*statements* are ceteris-paribus laws. It means that their nomological status is attributed to the specific physical context. The physical world is so complex that we must isolate some part of it and treat it as a nomological machine. Our example of physics or science as being the search for algorithmic compressions can be treated, in a new way, as a nomological machine in Cartwright's sense but presented in the context of mathematics. The model of the reflecting spheres in the description of complex systems and the fundamental role of causal relations as primitive concepts in Smolin and Unger's proposal, which produces laws as well as their mutability, might be considered as instances of contexts where nomological machines can be found, and where one might be confronted with fundamental, predictive limitations linked to uncomputability in various ways and contexts. In this abstract model, regularities or laws can emerge in two ways, namely as recurrence in the fundamental reflection process and as convergence to a specific limit set – in particular, the latter cannot be found in the usual representations of complex systems dynamics as network diagrams or graphs. We have also shown, in a new way, how Smolin and Unger's ideas of false universalism and false anachronism in modern cosmology can be viewed through Cartwright's ideas on nomological machines and that a correspondence exists with Tim Maudlin's concept of Fundamental Law of Temporal Evolution (FLOTE).

Furthermore, we showed that the standard cosmological itself model can be treated as a nomological machine. We used and modified Cartwright's concept to apply it to the cosmological model. Although it is not physical, the cosmological model can serve as a nomological machine thanks to its unique reference to the Universe, which is different from all the other models used in physics. However, it cannot be considered a model in the same manner as a planetary system model, as it does not correspond to any physical system that constitutes part of a greater whole. Instead, a cosmological model reflects everything – it is a model of that same whole. It can be said to enter a cognitive bond with the Universe, which we can only learn about through a model. SCM is a six-parameter model. As a nomological machine, it uses these basic parameters to arrive at other parameters that describe the Universe. These new parameters are independent and can be described using the basic six parameters and SCM equations.

For the models containing a cosmological component, classifying the possible evolutions according to the critical density is quite complicated. These models must first be reconstructed in the phase space to be organized. In this case, different values of the parameters of a model (density parameters) produce different types of evolution: (1) models that begin with an initial singularity to subsequently reach the maximal value of the scale factor, which then shrinks into a secondary singularity – called oscillatory models; (2) models that shrink until they reach the minimal size to expand into infinity – called bounce models subsequently; (3) models that expand from an initial singularity into Einstein's static universe; (4) models without an initial singularity that begins as Einstein's universe and evolves into infinity; and (5) models that begin with an initial singularity and evolve into infinity, where they become an asymptotically de Sitter universe, in which the scale factor is an exponential function of time. Study (Szydlowski & Tambor, 2017) shows that the evolution of cosmological models can be reduced to a dynamic Newtonian system. As with a planet with the Kepler potential, a particle moving in a field of potential mimics the universe's evolution, as expressed through changes in the scale factor, a(t). If the cosmological constant equals 0 (CDM model), the potential is precisely the Type 1/a Kepler potential. This construct is crucially essential for the ontological interpretation of the SCM as a nomological machine.

The machine delineates the evolutionary pathway of the model. It operates deterministically, and once the initial data are entered, it yields the result (relationship) a(t) or H(z). The machine constitutes an evolutionary scenario. As with planets, other fluids can be added to the model, beginning with the cosmological constant—Pawel Tambor in the monograph 'Standard Cosmological Model. A Methodological Study' worked out some of these specific aspects (Tambor 2020).

## Statements and Declarations

**Conflict of Interest**  On behalf of all authors, the corresponding author states that there is no conflict of interest. The authors have no relevant financial or non-financial interests to disclose. The authors have no competing interests to declare that are relevant to the content of this article. All authors certify that they have no affiliations with or involvement in any organization or entity with any financial interest or non-financial interest in the subject matter or materials discussed in this manuscript.

**Dr. Yves Gaspar** obtained in 2002 a Ph.D. in theoretical cosmology from the University of Cambridge (UK), with Prof. John Barrow as supervisor. He is lecturer at the Catholic University of the Sacred Heart in Italy, and is currently Visiting Scholar at St. Edmund's College of the University of Cambridge, where he has carried out research on the complexity of the field equations of general relativity and cosmology, on uncomputability and on the possible philosophical and theological implications.

**Pawel Tambor, Ph.D,** is a lecturer and researcher at The John Paul II Catholic University of Lublin, Poland, and assistant professor at Jan Kochanowski University of Kielce, Poland. His fields of research and specialty are philosophy of science, philosophy of modern cosmology, modeling in science and the relations between science, philosophy and theology.